\documentstyle[aps,preprint]{revtex}
\input epsf
\global\firstfigfalse  % kludge to bypass foolishness of revtex to place

\tighten

\begin{document}
\pagestyle{empty}

\preprint{
\begin{minipage}[t]{3in}
%\begin{flushright} IAS--TH--00/18,  NSF-ITP-00-16 \\
%hep-th/0003136 \\
%March 2000
%\end{flushright}
\end{minipage}
}

\title{New generalized Chaplygin gas as a scheme for unification of
dark energy and dark matter}

\author{Xin Zhang$^{1,2}$, Feng-Quan Wu$^{1,2}$, and Jingfei Zhang$^{3}$\\\bigskip}
\address{1 CCAST (World Laboratory), P.O.Box 8730, Beijing 100080, People's Republic of
China\\ 2 Institute of High Energy Physics, Chinese Academy of
Sciences, P.O.Box 918(4), Beijing 100049, People's Republic of China\\
3 Department of Physics, Liaoning Normal University, Dalian
116029, People's Republic of China}

\maketitle

\begin{abstract}

We propose in this paper a new model for describing the
unification of dark energy and dark matter. This new model is a
further generalization of the generalized Chaplygin gas (GCG)
model, thus dubbed new generalized Chaplygin gas (NGCG) model. The
equation of state of the NGCG is given by $p = -
\tilde{A}(a)/\rho^{\alpha}$, where $a$ is the scale factor and
$\tilde{A}(a)=-w_XAa^{-3(1+w_X)(1+\alpha)}$. We show that the NGCG
model is totally dual to an interacting XCDM parametrization
scenario, in which the interaction between dark energy and dark
matter is characterized by the constant $\alpha$. We discuss the
cosmological consequences led by such an unified dark sectors
model. Furthermore, we perform a statefinder analysis on this
scenario and show the discrimination between this scenario and
other dark energy models. Finally, a combined analysis of the data
of Type Ia supernovae, cosmic microwave background, and large
scale structure provides a fairly tight constraint on the
parameters of the NGCG model.

\vskip .4cm

%\noindent PACS: 04.70.Dy;02.20.Uw;97.60.Lf

\vskip .2cm

%\noindent Keywords: Schwarzschild black hole; evaporation;
%$q$-deformation; noncommutative spacetime.

\end{abstract}

\newpage
\pagestyle{plain} \narrowtext \baselineskip=18pt

\setcounter{footnote}{0}

\section{Introduction}

Recent observations of Type Ia supernovae (SNe Ia) \cite{sn}
indicate that the expansion of the Universe is accelerating at the
present time. These results, when combined with the observations
of cosmic microwave background (CMB) \cite{wmap} and large scale
structure (LSS) \cite{sdss}, strongly suggest that the Universe is
spatially flat and dominated by an exotic component with large
negative pressure, referred to as dark energy \cite{de}. The first
year result of the Wilkinson Microwave Anisotropy Probe (WMAP)
shows that dark energy occupies about $73\%$ of the energy of our
Universe, and dark matter about $23\%$. The usual baryon matter
which can be described by our known particle theory occupies only
about $4\%$ of the total energy of the Universe. Although we can
affirm that the ultimate fate of the Universe is determined by the
feature of dark energy, the nature of dark energy as well as its
cosmological origin remain enigmatic at present. So far the
confirmed information about dark energy is still limited and can
be roughly summarized as the following several items: it is a kind
of exotic energy form with sufficiently large negative pressure
such that drives the Universe to undergo a period of accelerating
expansion currently; it is spatially homogeneous and
non-clustering; and it is in small part at the early times while
dominates the Universe very recently.

The investigation of the nature of dark energy is an important
mission in the modern cosmology. Much work has been done on this
issue, and there is still a long way to go. Currently, the
preferred candidates of dark energy are vacuum energy (or
cosmological constant) and dynamical fields. The simplest form of
dark energy is the cosmological constant $\Lambda$. A tiny
positive cosmological constant which can naturally explain the
current acceleration would encounter many theoretical problems,
such as the ``fine-tuning'' problem and the ``coincidence''
problem. Another possible form of dark energy is provided by
scalar fields. Dark energy can be attributed to the dynamics of a
scalar field $\phi$, called quintessence \cite{quin}, which
convincingly realize the present accelerated expansion of the
Universe by evolving slowly down its potential $V(\phi)$. The
tracker version quintessence is to some extent likely to resolve
the coincidence problem. It should also be pointed out that the
coupled quintessence models \cite{couple} provide this problem
with a more natural solution. However, for quintessence models
with flat potentials, the quintessence field has to be nearly
massless and one thus expects radiative corrections to destabilize
the ratio between this mass and the other known scales of physics.
In addition, for the cosmological constant and many quintessence
models, the event horizon would lead to a potential
incompatibility with the string theory.

Other designs on dark energy roughly include k-essence
\cite{kess}, quiessence (or ``X-matter'') \cite{xmatter}, brane
world \cite{brane}, tachyon \cite{tachyon}, generalized Chaplygin
gas (GCG) \cite{chaplygin,gcgtest}, holographic dark energy
\cite{holography,holoSN}, and so forth. The quiessence or X-matter
component is simply characterized by a constant, non-positive
equation of state $w_X$ ($w_X<-1/3$ is a necessary condition to
make the Universe accelerate). In general, in a
Friedmann-Robertson-Walker (FRW) background with the presence of
cold dark matter (CDM), an arbitrary but constant $w_X$ for dark
energy from the range $(-1,0)$ can be achieved by using a scalar
field with a hyperbolic sine potential \cite{xmatter}. It may be
noted that in principle the value of $w_X$ may be even less than
$-1$. In fact, by fitting the SNe Ia data in the framework of XCDM
(X-matter with CDM), the hint for $w_X<-1$ has been found. Indeed,
a study of high-$z$ SNe Ia \cite{Knop} finds that the equation of
state of dark energy has a $99\%$ probability of being $<-1$ if no
priors are placed on $\Omega_m^0$. When these SNe results are
combined with CMB and 2dFGRS the $95\%$ confidence limits on an
unevolving equation of state are $-1.46<w_X<-0.78$
\cite{Knop,Riess} which is consistent with estimates made by other
groups \cite{wmap,sdss}. The possibility of $w_X<-1$ has provoked
lots of investigations on the phantom dark energy \cite{phantom}.
The remarkable feature of the phantom model is that the Universe
will end its life with a ``Big Rip'' (future singularity) within a
finite time \cite{bigrip}. On the other hand, we concern here
another interesting proposal on dark energy, i.e. the dark energy
component might be explained by a background fluid with an exotic
equation of state, the generalized Chaplygin gas model
\cite{chaplygin}. The striking feature of this model is that it
allows for a unification of dark energy and dark matter. This
point can be easily seen from the fact that the GCG behaves as a
dust-like matter at early times and behaves like a cosmological
constant at late stage. This dual role is at the heart of the
surprising properties of the GCG model. Moreover, the GCG model
has been successfully confronted with various phenomenological
tests involving SNe Ia data, CMB peak locations, gravitational
lensing and other observational data \cite{gcgtest}. It is
remarkable that the GCG equation of state has a well defined
connection with string and brane theories \cite{gcgbrane}, and
this gas is the only gas known to admit a supersymmetric
generalization \cite{susy}.

In addition, it should be pointed out that the GCG model can be
portrayed as a picture that cosmological constant type dark energy
interacts with cold dark matter. However, since the equation of
state of dark energy still cannot be determined exactly, the
observational data show $w_X$ is in the range of $(-1.46, -0.78)$,
the GCG model should naturally be generalized to accommodate any
possible X-type dark energy with constant $w_X$. Therefore, we
propose here a new generalized Chaplygin gas (NGCG) scenario as a
scheme for unification of X-type dark energy and dark matter. The
feature of this new model should also be exhibited in that dark
sectors are uniformly described by an exotic background fluid and
this new gas behaves as a dust-like matter at early times and as a
X-type dark energy at late times. We will show in this paper that
this model is a kind of interacting XCDM model, and constrain the
parameters of this model by using observational data.

This paper is organized as follows. In Section 2, we introduce the
extension version of the generalized Chaplygin gas, namely the
NGCG model, to describe the unification of dark energy and dark
matter, and demonstrate that the NGCG actually is a kind of
interacting XCDM system. In Section 3, we analyze the NGCG model
by means of the statefinder parameters. In Section 4, we constrain
the parameters of the NGCG model using the SNe Ia, CMB, and LSS
data. We give concluding remarks in the final section.

\section{The NGCG scenario and the interacting XCDM parametrization}

%\subsection{The NGCG scenario}

In this section we introduce the NGCG model. In the framework of
FRW cosmology, considering an exotic background fluid, the NGCG,
described by the equation of state
\begin{equation}
p_{\rm Ch} = - {\tilde{A}(a) \over \rho_{\rm Ch}^\alpha}~,
\label{eqstate}
\end{equation}
where $\alpha$ is a real number and $\tilde{A}(a)$ is a function
depends upon the scale factor of the Universe, $a$. We might
expect that this exotic background fluid smoothly interpolates
between a dust dominated phase $\rho\sim a^{-3}$ and a dark energy
dominated phase $\rho\sim a^{-3(1+w_X)}$ where $w_X$ is a constant
and should be taken as any possible value in the range
$(-1.46,-0.78)$. It can be expected that the energy density of the
NGCG should be elegantly expressed as
\begin{equation}
\rho_{\rm Ch}=  \left[A a^{-3(1+w_X)(1+\alpha)} + {B a^{-3 (1 +
\alpha)}}\right]^{1 \over 1 + \alpha}~.\label{chrho}
\end{equation}
The derivation of the Eq. (\ref{chrho}) should be the consequence
of substituting the equation of state Eq. (\ref{eqstate}) into the
energy conservation equation of the NGCG for an homogeneous and
isotropic spacetime, this requires the function $\tilde{A}(a)$ to
be of the form
\begin{equation}
\tilde{A}(a)=-w_XAa^{-3(1+w_X)(1+\alpha)}~,\label{A}
\end{equation}
where $A$ is a positive constant, and the other positive constant
$B$ appears in Eq. (\ref{chrho}) an integration constant. One can
see explicitly that this model recovers the GCG model as the
equation-of-state parameter $w_X$ taken to be $-1$, and an
ordinary $X$CDM model can be reproduced by taking the parameter
$\alpha$ to be zero. The parameter $\alpha$ is called {\it
interaction parameter} of the model as will be shown below.

The NGCG scenario involves an interacting XCDM picture. For
showing this, we first decompose the NGCG fluid into two
components, one is the dark energy component, and the other is the
dark matter component,
\begin{equation}
\rho_{\rm Ch}=\rho_X+\rho_{dm}~.
\end{equation}
Note that the pressure of the NGCG fluid is provided only by the
dark energy component, namely $p_{\rm Ch}=p_X$. Therefore, the
energy density of the dark energy ingredient can be given
\begin{equation} \rho_X={p_{\rm Ch}\over w_X}={A a^{-3(1+w_X)(1+\alpha)}\over
[Aa^{-3(1+w_X)(1+\alpha)}+Ba^{-3(1+\alpha)}]^{\alpha\over
1+\alpha}}~~,\label{rhox}\end{equation} and then the energy
density of the dark matter component can also be obtained
\begin{equation}
\rho_{dm}={Ba^{-3(1+\alpha)}\over
[Aa^{-3(1+w_X)(1+\alpha)}+Ba^{-3(1+\alpha)}]^{\alpha\over
1+\alpha}}~~.\label{rhodm}\end{equation} From these expressions
one obtains the scaling behavior of the energy densities
\begin{equation} {\rho_{dm}\over\rho_X}={B\over
A}a^{3w_X(1+\alpha)}~~.\label{scaling}\end{equation} We see
explicitly from it that there must exist an energy flow between
dark matter and dark energy provided that $\alpha\neq 0$. When
$\alpha>0$, the transfer direction of the energy flow is from dark
matter to dark energy; when $\alpha<0$, just the reverse.
Therefore, it is clear that the parameter $\alpha$ characterizes
the interaction between dark energy and dark matter. This is the
reason for that we call $\alpha$ the interaction parameter.

The parameters $A$ and $B$ can be expressed in terms of current
cosmological observables. From Eq.(\ref{chrho}), it is easy to get
\begin{equation}
A+B=\rho_{\rm Ch0}^\eta~,
\end{equation}
where $\eta=1+\alpha$ is used to characterize the interaction for
simplicity, thus we have
\begin{equation}
A=\rho_{\rm Ch0}^\eta A_s~,~~~~B=\rho_{\rm Ch0}^\eta
(1-A_s)~,\label{As}
\end{equation} where $A_s$ is a dimensionless parameter.
Using Eqs. (\ref{scaling}) and (\ref{As}), one gets
\begin{equation}
A_s={\rho_{X0}\over \rho_{X0}+\rho_{dm0}}={\Omega_X^0\over
1-\Omega_{b}^0}~,
\end{equation}
where the second equality stands for the cosmological model
involving the baryon matter component. We have assumed here that
the space of the Universe is flat. Hence, the NGCG energy density
can be expressed as
\begin{equation}
\rho_{\rm Ch}=\rho_{\rm Ch0}
a^{-3}[1-A_s(1-a^{-3w_X\eta})]^{1/\eta}~.\label{ngcg}
\end{equation}
Making use of Eqs. (\ref{chrho}), (\ref{rhox}), (\ref{rhodm}), and
(\ref{ngcg}), the energy densities of dark energy and dark matter
can be re-expressed as
\begin{equation}
\rho_X=\rho_{X0} a^{-3(1+w_X\eta)}\left[1-{\Omega_X^0\over
1-\Omega_{b}^0}(1-a^{-3w_X\eta})\right]^{{1\over\eta}-1}~,\label{de}
\end{equation}
\begin{equation}
\rho_{dm}=\rho_{dm0} a^{-3}\left[1-{\Omega_X^0\over
1-\Omega_{b}^0}(1-a^{-3w_X\eta})\right]^{{1\over\eta}-1}~.\label{dm}
\end{equation}

The whole NGCG fluid satisfies the energy conservation, but dark
energy and dark matter components do not obey the energy
conservation separately; they interact with each other. We depict
this interaction through an energy exchange term $Q$. The
equations of motion for dark energy and dark matter can be written
as
\begin{equation}
\dot{\rho}_X+3H(1+w_X)\rho_X=Q~,
\end{equation}
\begin{equation}
\dot{\rho}_{dm}+3H\rho_{dm}=-Q~,
\end{equation}
where dot denotes a derivative with respect to time $t$, and
$H=\dot{a}/a$ represents the Hubble parameter. For convenience we
define the effective equations of state for dark energy and dark
matter through the parameters
\begin{equation}
w_X^{(e)}=w_X-{Q\over 3H\rho_X}~,
\end{equation}
\begin{equation}
w_{dm}^{(e)}={Q\over 3H\rho_{dm}}~.
\end{equation}
According to the definition of the effective equations of state,
the equations of motion for dark energy and dark matter can be
re-expressed into forms of energy conservation,
\begin{equation}
\dot{\rho}_X+3H(1+w_X^{(e)})\rho_X=0~,
\end{equation}
\begin{equation}
\dot{\rho}_{dm}+3H(1+w_{dm}^{(e)})\rho_{dm}=0~.
\end{equation}
By means of the concrete forms of dark energy and dark matter in
NGCG scenario, Eqs. (\ref{de}) and (\ref{dm}), one can obtain
\begin{equation}
w_X^{(e)}=w_X+{(\eta-1)w_X(1-\Omega_X^0-\Omega_b^0)a^{3w_X\eta}\over
\Omega_X^0+(1-\Omega_X^0-\Omega_b^0)a^{3w_X\eta}}~,\label{wXeff}
\end{equation}
\begin{equation}
w_{dm}^{(e)}=-{(\eta-1)w_X\Omega_X^0\over
\Omega_X^0+(1-\Omega_X^0-\Omega_b^0)a^{3w_X\eta}}~.
\end{equation}

Now we switch to discuss the cosmological evolution. Consider a
spatially flat FRW Universe with baryon matter component
$\rho_{b}$ and NGCG fluid $\rho_{\rm Ch}$, the Friedmann equation
reads
\begin{equation}
3M_P^2H^2=\rho_{\rm Ch}+\rho_{b}~,
\end{equation}
where $M_P$ is the reduced Planck mass. The Friedmann equation can
also be expressed as
\begin{equation}
H(a)=H_0E(a)~,
\end{equation}
where
\begin{equation}
E(a)=\left\{(1-\Omega_b^0)a^{-3}\left[1-{\Omega_X^0\over
1-\Omega_b^0}(1-a^{-3w_X\eta})\right]^{1/\eta}+\Omega_b^0
a^{-3}\right\}^{1/2}~.
\end{equation}
Then, the fractional energy densities of various components can be
easily obtained
\begin{equation}
\Omega_X=\Omega_X^0
E^{-2}a^{-3(1+w_X\eta)}\left[1-{\Omega_X^0\over 1-\Omega_{
b}^0}(1-a^{-3w_X\eta})\right]^{{1\over\eta}-1}~,\label{OmegaX}
\end{equation}
\begin{equation}
\Omega_{dm}=(1-\Omega_X^0-\Omega_{b}^0)E^{-2}a^{-3}\left[1-{\Omega_X^0\over
1-\Omega_{b}^0}(1-a^{-3w_X\eta})\right]^{{1\over\eta}-1}~,
\end{equation}
\begin{equation}
\Omega_{b}=\Omega_{b}^0E^{-2}a^{-3}~.
\end{equation}

So far we see clearly that the NGCG model is totally dual to a
coupled dark energy scenario \cite{IXCDM,Cai}, namely an
interacting XCDM parametrization. It is remarkable that the
interaction between dark energy and dark matter can be interpreted
as arising from the time variation of the mass of dark matter
particles. The GCG model is a special case in the NGCG model
corresponding to $w_X=-1$, thus the GCG model is actually an
interacting $\Lambda$CDM model. Fig.1 and Fig.2 illustrate
examples of the density evolution in the NGCG model. The current
density parameters used in the plots are $\Omega_{dm}^0=0.25$,
$\Omega_X^0=0.7$, and $\Omega_b^0=0.05$. In Fig.1, we show the
cases having the common equation-of-state parameter $w_X=-1.2$,
while the interaction parameters $\alpha$ are taken to be $0$,
$0.5$, and $0.8$, respectively. Note that the $\alpha=0$ case
corresponds to a normal phantom model with constant $w_X$. In this
example we see the role the interaction parameter $\alpha$ plays
in the model. The transfer energy flows from dark matter to dark
energy when $\alpha>0$; the larger $\alpha$ leads to the stronger
energy flow; density of baryon component is also affected
evidently by the interaction between dark energy and dark matter.
In Fig.2, we depict the cases with common interaction parameter
$\alpha=0.5$, and the equation-of-state parameters $w_X$ are taken
to be $-1$, $-0.8$, and $-1.2$, respectively. Here $w_X=-1$ case
corresponds exactly to the GCG model. The effect of the parameter
$w_X$ in the NGCG scenario is also evident as we see in this
example.

%%%%%%%%%%%%%%%%%%%%%%%%%%%%%%%%%%%%%%%%%%%%%%%%%%%%%%%%%%%%%%%%$
\vskip.8cm
\begin{figure}
\begin{center}
\leavevmode \epsfbox{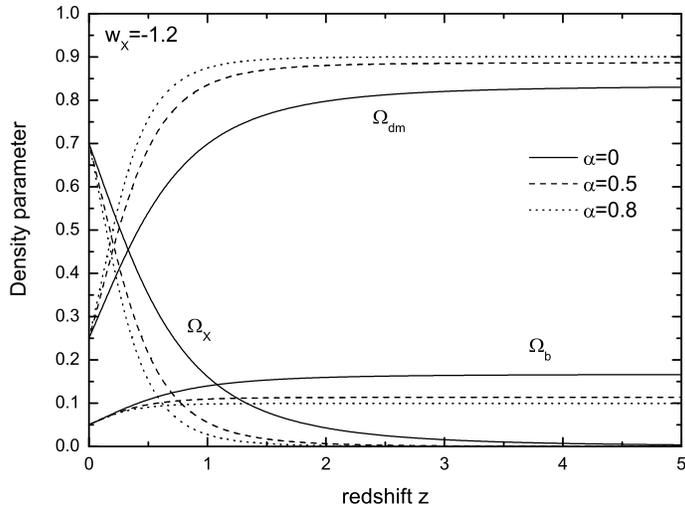} \caption[]{The evolution of the
density parameters for various components $\Omega_X$,
$\Omega_{dm}$, and $\Omega_b$. Note that $\Omega_{\rm
Ch}=\Omega_X+\Omega_{dm}$. The current density parameters used in
the plot are $\Omega_{dm}^0=0.25$, $\Omega_X^0=0.7$, and
$\Omega_b^0=0.05$. In this case, we fix $w_X$ and vary $\alpha$. }
\end{center}
\end{figure}
%%%%%%%%%%%%%%%%%%%%%%%%%%%%%%%%%%%%%%%%%%%%%%%%%%%%%%%%%%%%%%%%%

%%%%%%%%%%%%%%%%%%%%%%%%%%%%%%%%%%%%%%%%%%%%%%%%%%%%%%%%%%%%%%%%$
\vskip.8cm
\begin{figure}
\begin{center}
\leavevmode \epsfbox{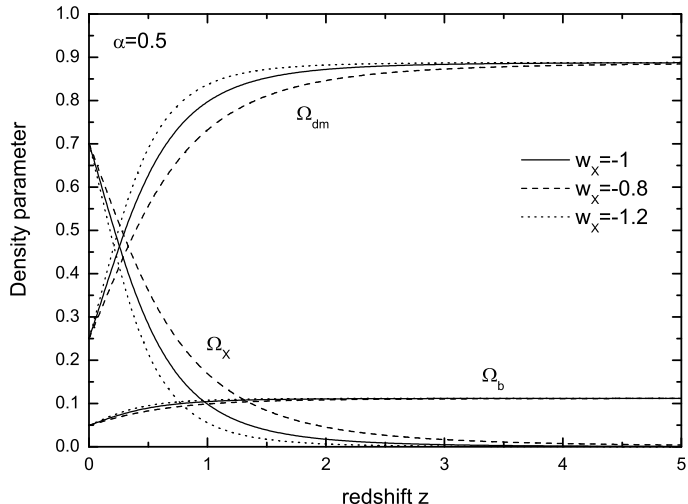} \caption[]{The evolution of the
density parameters for various components $\Omega_X$,
$\Omega_{dm}$, and $\Omega_b$. Note that $\Omega_{\rm
Ch}=\Omega_X+\Omega_{dm}$. The current density parameters used in
the plot are $\Omega_{dm}^0=0.25$, $\Omega_X^0=0.7$, and
$\Omega_b^0=0.05$. In this case, we fix $\alpha$ and vary $w_X$.}
\end{center}
\end{figure}
%%%%%%%%%%%%%%%%%%%%%%%%%%%%%%%%%%%%%%%%%%%%%%%%%%%%%%%%%%%%%%%%%

%\subsection{Cosmological consequences}

Let us now discuss the cosmological consequences led by the NGCG
model and compare the cosmological quantities in the NGCG
cosmology with those of some special cases such as $\Lambda$CDM
and GCG. Firstly, we regard the Hubble parameter $H$ which
evaluates the expansion rate of the Universe. In Fig.3 we plot the
Hubble parameter of the NGCG model in units of $H_{\rm \Lambda
CDM}$ as a function of redshift $z$ range from 0 to 5. The current
density parameters used in the plot of Fig.3 are the same as used
in Figs.1 and 2. The model parameters are divided into two groups,
$\alpha=0$ and $\alpha=0.5$, both including $w_X=-1$, $-0.8$, and
$-1.2$. It can be seen from Fig.3 that the NGCG model degenerates
to the XCDM when the parameter $\alpha$ takes 0; the cases of
$w_X>-1$ and $w_X<-1$ make $H$ larger than and less than $H_{\rm
\Lambda CDM}$, respectively, during the cosmological evolution.
The introducing of the interaction parameter $\alpha$ makes $H$ be
larger than $H_{\rm \Lambda CDM}$ evidently at early times; while
it is interesting to see that the value of $H/H_{\rm \Lambda CDM}$
can cross 1 in the case of $w_X<-1$ in recent period. The
acceleration of the Universe is evaluated by the deceleration
parameter $q=-\ddot{a}/aH^2$. Omitting the radiation component,
the deceleration parameter can be expressed as
\begin{equation}
q={1\over 2}+{3\over 2}w_X\Omega_X~,
\end{equation}
where $\Omega_X$ is given by (\ref{OmegaX}). The evolution of the
deceleration parameter $q$ is depicted in Fig.3 for selected
parameter sets. The current density parameters are taken to be the
same as above figures. The influence coming from the interaction
$\alpha$ and equation of state of dark energy $w_X$ can be seen
clearly in this figure. We notice that a positive $\alpha$ makes
the redshift of acceleration/deceleration transition ($q(z_T)=0$)
shift to a smaller value; while the values of $z_T$ are nearly
degenerate under the same $\alpha$ as shown in this example.

%%%%%%%%%%%%%%%%%%%%%%%%%%%%%%%%%%%%%%%%%%%%%%%%%%%%%%%%%%%%%%%%$
\vskip.8cm
\begin{figure}
\begin{center}
\leavevmode \epsfbox{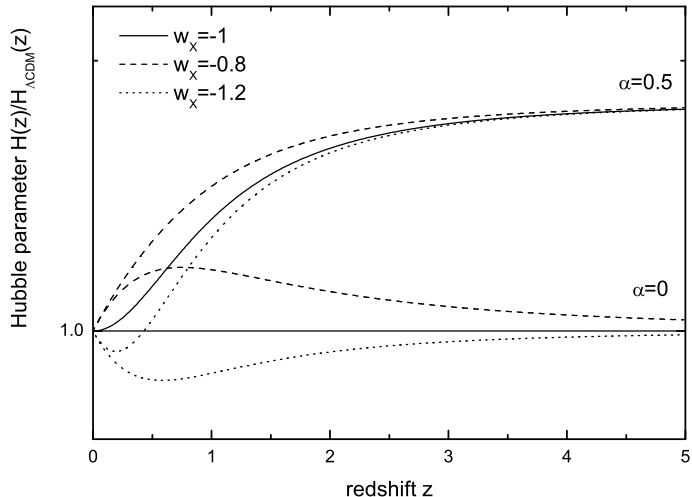} \caption[]{The evolution of the Hubble
parameter in units of $H_{\rm \Lambda CDM}(z)$. The current
density parameters are taken to be $\Omega_{dm}^0=0.25$,
$\Omega_X^0=0.7$, and $\Omega_b^0=0.05$.}
\end{center}
\end{figure}
%%%%%%%%%%%%%%%%%%%%%%%%%%%%%%%%%%%%%%%%%%%%%%%%%%%%%%%%%%%%%%%%%

%%%%%%%%%%%%%%%%%%%%%%%%%%%%%%%%%%%%%%%%%%%%%%%%%%%%%%%%%%%%%%%%$
\vskip.8cm
\begin{figure}
\begin{center}
\leavevmode \epsfbox{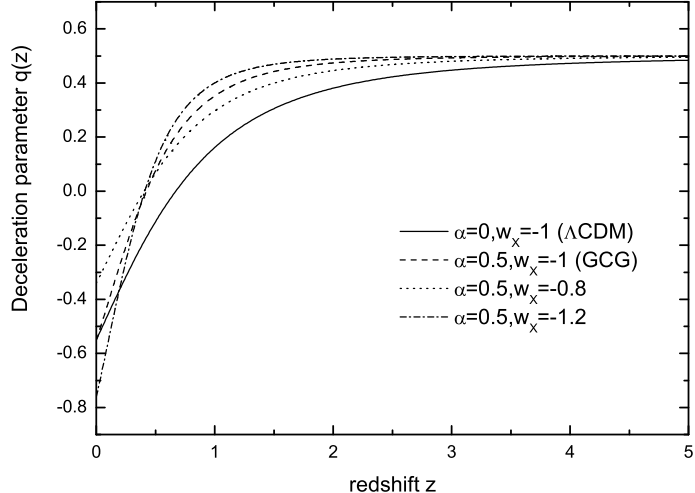} \caption[]{The evolution of the
deceleration parameter $q(z)$. The current density parameters are
taken to be $\Omega_{dm}^0=0.25$, $\Omega_X^0=0.7$, and
$\Omega_b^0=0.05$.}
\end{center}
\end{figure}
%%%%%%%%%%%%%%%%%%%%%%%%%%%%%%%%%%%%%%%%%%%%%%%%%%%%%%%%%%%%%%%%%

\section{Statefinder diagnostic}

Since more and more dark energy models have been constructed for
interpreting or describing the cosmic acceleration, the problem of
discriminating between the various contenders becomes very
important. In order to be able to differentiate between those
competing cosmological scenarios involving dark energy, a
sensitive and robust diagnostic for dark energy models is a must.
For this purpose a diagnostic proposal that makes use of parameter
pair $\{r,s\}$, the so-called ``statefinder'', was introduced by
Sahni et al. \cite{sahni}. The statefinder probes the expansion
dynamics of the Universe through higher derivatives of the scale
factor $\stackrel{...}{a}$ and is a natural companion to the
deceleration parameter $q$ which depends upon $\ddot a$. The
statefinder pair $\{r,s\}$ is defined as follows
\begin{equation}
r\equiv
\frac{\stackrel{...}{a}}{aH^3},~~~~s\equiv\frac{r-1}{3(q-1/2)}~.\label{rs}
\end{equation} The statefinder
is a ``geometrical'' diagnostic in the sense that it depends upon
the scale factor and hence upon the metric describing space-time.

Trajectories in the $s-r$ plane corresponding to different
cosmological models exhibit qualitatively different behaviors. The
spatially flat $\Lambda$CDM scenario corresponds to a fixed point
in the diagram
\begin{equation}
\{s,r\}\bigg\vert_{\rm \Lambda CDM} = \{ 0,1\} ~.\label{lcdm}
\end{equation}
Departure of a given dark energy model from this fixed point
provides a good way of establishing the ``distance'' of this model
from $\Lambda$CDM \cite{sahni,alam}. As demonstrated in Refs.
\cite{sahni,alam,quintomsr,gorini,holosr,zimdahl,zx} the
statefinder can successfully differentiate between a wide variety
of dark energy models including the cosmological constant,
quintessence, quintom, the Chaplygin gas, braneworld models,
holographic dark energy and interacting dark energy models. We can
clearly identify the ``distance'' from a given dark energy model
to the $\Lambda$CDM scenario by using the $r(s)$ evolution
diagram.

The current location of the parameters $s$ and $r$ in these
diagrams can be calculated in models, and on the other hand it can
also be extracted from data coming from SNAP (SuperNovae
Acceleration Probe) type experiments \cite{sahni,alam}. Therefore,
the statefinder diagnostic combined with future SNAP observations
may possibly be used to discriminate between different dark energy
models. For example, as shown in Ref. \cite{alam}, by carrying out
a maximum likelihood analysis which combines the statefinder
diagnostic with realistic expectations from the SNAP experiment,
the averaged-over-redshift statefinder pair $\{\bar{s}, \bar{r}\}$
is convincingly demonstrated to be useful diagnostic tool in
successfully differentiating between the cosmological constant and
dynamical models of dark energy. In this section we apply the
statefinder diagnostic to the NGCG model.

In what follows we will calculate the rangefinder parameters for
the NGCG model and plot the evolution trajectories of the model in
the statefinder parameter-plane. The statefinder parameters can be
expressed in terms of the total energy density $\rho$ and the
total pressure $p$ in the Universe:
\begin{equation}
r=1+{9(\rho+p)\over
2\rho}{\dot{p}\over\dot{\rho}}~,~~~~s={(\rho+p)\over
p}{\dot{p}\over\dot{\rho}}~.
\end{equation} The total energy of the Universe is
conserved, so we have $\dot{\rho}=-3H(\rho+p).$ Since the dust
matter does not have pressure, the total pressure of the cosmic
fluids is provided only by dark energy component,
$p=p_X=w_X\rho_X$. Then making use of $\dot{\rho}=-3H(\rho+p)$ and
$\dot{\rho}_X=-3H(1+w_X^{(e)})\rho_X$, we can get the concrete
expression of the statefinder parameters
\begin{equation}
r=1+{9\over 2}w_X\Omega_X(1+w_X^{(e)})~,~~~~s=1+w_X^{(e)}~.
\end{equation} Here $w_X^{(e)}$ and $\Omega_X$ are given by (\ref{wXeff})
and (\ref{OmegaX}), respectively. Though the relationship between
statefinder parameters $r$ and $s$, namely the function $r(s)$,
might be derived analytically in principle, we do not give the
expression here due to the complexity of the formula. Making the
redshift $z={1/a}-1$ vary in an enough large range involving far
future and far past, e.g. from $-1$ to 5, one can easily get the
evolution trajectories in the statefinder $s-r$ plane of this
model. Selected curves of $r(s)$ are plotted in Fig.5 and Fig.6.
In Fig.5, we fix $\alpha=0.5$ and vary $w_X$ as $-0.8$, $-1$, and
$-1.2$, respectively. In Fig.6, we fix $w_X=-1.2$ and vary
$\alpha$ as 0, $\pm0.2$, $\pm0.5$, and $\pm0.8$, respectively.
Other parameters are taken as the same as previous figures. In
these two figures, dots locate the today's values of the
statefinder parameters $(s_0, r_0)$ and arrows denote the
evolution directions of the statefinder trajectories $r(s)$. The
$\Lambda$CDM model locates at $(0, 1)$ in the $s-r$ plane also
denoted as a dot.

%%%%%%%%%%%%%%%%%%%%%%%%%%%%%%%%%%%%%%%%%%%%%%%%%%%%%%%%%%%%%%%%$
\vskip.8cm
\begin{figure}
\begin{center}
\leavevmode \epsfbox{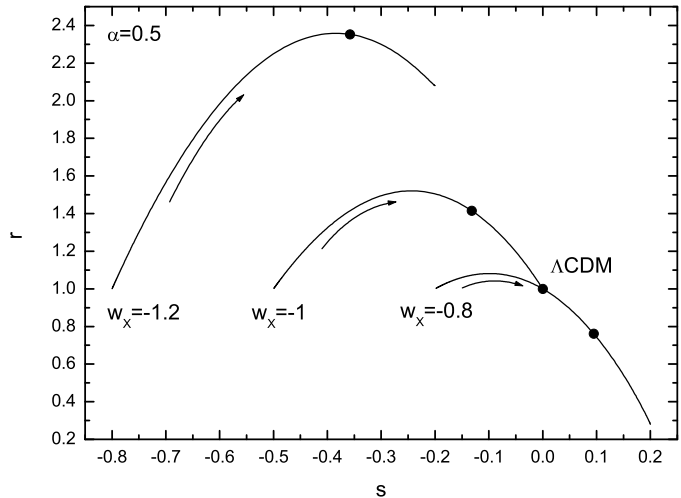} \caption[]{The statefinder $r(s)$
evolution diagram. Dots locate the today's values of the
statefinder parameters $(s_0, r_0)$ and arrows denote the
evolution directions of the statefinder trajectories $r(s)$. The
$\Lambda$CDM model locates at the fixed point $(0, 1)$. In this
case, we fix $\alpha=0.5$ and vary $w_X$ as $-0.8$, $-1$, and
$-1.2$, respectively.}
\end{center}
\end{figure}
%%%%%%%%%%%%%%%%%%%%%%%%%%%%%%%%%%%%%%%%%%%%%%%%%%%%%%%%%%%%%%%%%

%%%%%%%%%%%%%%%%%%%%%%%%%%%%%%%%%%%%%%%%%%%%%%%%%%%%%%%%%%%%%%%%$
\vskip.8cm
\begin{figure}
\begin{center}
\leavevmode \epsfbox{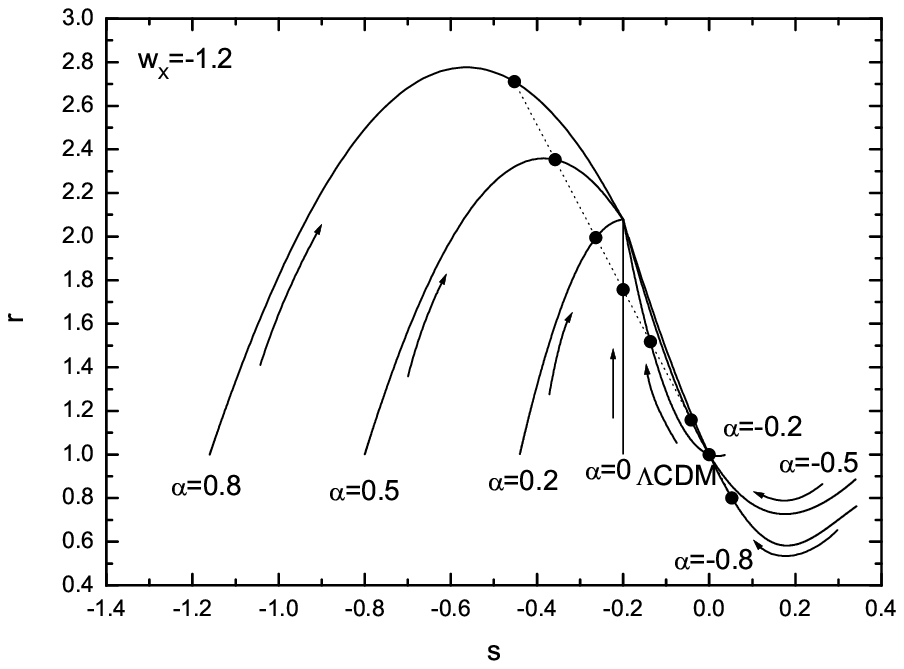} \caption[]{The statefinder $r(s)$
evolution diagram. Dots locate the today's values of the
statefinder parameters $(s_0, r_0)$ and arrows denote the
evolution directions of the statefinder trajectories $r(s)$. The
$\Lambda$CDM model locates at the fixed point $(0, 1)$. In this
case, we fix $w_X=-1.2$ and vary $\alpha$ as 0, $\pm0.2$,
$\pm0.5$, and $\pm0.8$, respectively.}
\end{center}
\end{figure}
%%%%%%%%%%%%%%%%%%%%%%%%%%%%%%%%%%%%%%%%%%%%%%%%%%%%%%%%%%%%%%%%%

The statefinder diagnostic can discriminate between various dark
energy models effectively. Different cosmological models involving
dark energy exhibit qualitatively different evolution trajectories
in the $s-r$ plane. For example, the $\Lambda$CDM scenario
corresponds to the fixed point $s=0,~r=1$ as shown in
(\ref{lcdm}), and the SCDM (standard cold dark matter) scenario
corresponds to the point $s=1,~r=1$. For the ``quiessence" (XCDM)
models, the trajectories are some vertical segments, i.e. $r$
decreases monotonically from 1 to $1+{9\over 2}w_X(1+w_X)$ while
$s$ remains constant at $1+w_X$ \cite{sahni,alam}. The
quintessence (inverse power law) tracker models have typical
trajectories similar to arcs of an upward parabola lying in the
regions $s>0,~r<1$ \cite{sahni,alam}. The holographic dark energy
scenario ($c=1$ case), as shown in \cite{holosr}, commences its
evolution from $s=2/3,~r=1$, through an arc segment, and ends it
at the $\Lambda$CDM fixed point ($s=0,~r=1$) in the future. The
coupled quintessence models and Quintom models exhibit more
complicated trajectories as shown in Refs. \cite{quintomsr,zx}.
Now from the Figs. 5 and 6 of this paper, we can see the
statefinder trajectories of the NGCG model. In Fig.5 we see the
cases under the fixed $\alpha$, where the GCG model ($w_X=-1$)
exhibits a complete downward parabola, while the general cases
($w_X\neq -1$) correspond to some broken parabolas. The
statefinder trajectory commences its evolution from
$s=1+(1+\alpha)w_X,~r=1$ at $t\rightarrow 0$ to
$s=1+w_X,~r=1+{9\over 2}w_X(1+w_X)$ at $t\rightarrow \infty$,
through the curve. Today's statefinder point locates at
$s_0=1+w^{(e)}_{X0}$, $r_0=1+{9\over
2}w_X\Omega_X^0(1+w^{(e)}_{X0})$, where $w^{(e)}_{X0}=w_X+\alpha
w_X(1-\Omega_X^0-\Omega_b^0)/(1-\Omega_b^0)$. The ``distance''
from the NGCG model to the $\Lambda$CDM scenario can be measured
directly in the statefinder plane. Note that under a positive
$\alpha$, the cases of $w_X<-1$ never arrive at the $\Lambda$CDM
fixed point; the GCG case ($w_X=-1$) ends at the $\Lambda$CDM
fixed point; while the cases of $w_X>-1$ have passed through this
fixed point. Fig.6 displays the cases under a fixed $w_X$. We show
here $w_X=-1.2$, a phantom. Trajectories correspond to zero,
positive, as well as negative values of $\alpha$ are all displayed
in this diagram to depict a complete statefinder diagnostic. It is
interesting to see that the trajectories can pass through the
$\Lambda$CDM fixed point under a phantom case when $\alpha<0$.
This is because a negative $\alpha$ makes dark energy component
transfer energy flow to dark matter component. We notice that the
normal phantom case ($\alpha=0$) evolves its trajectory along a
vertical segment. Comparing with the quiessence case
\cite{sahni,alam}, the phantom case reposes on the left of
$\Lambda$CDM point, namely the region $s<0,~r>1$, and evolves
upwards; while the quiessence case reposes on the right of the
$\Lambda$CDM point, namely the region $s>0,~r<1$, and evolves
downwards. Interestingly, under a fixed $w_X$, the present
statefinder points as well as the $\Lambda$CDM fixed point locate
on a straight line. This is because when $w_X$ is fixed, the
relationship between $r_0$ and $s_0$ is linear, $r_0=1+{9\over
2}w_X\Omega_X^0 s_0$.

\section{Observational constraints from SNe Ia, CMB, and LSS data}

In this section we will derive the constraints on the NGCG model
from current available observational data. It should be mentioned
that the interacting XCDM parametrization scenario has been tested
by the recent Type Ia supernovae data \cite{IXCDM}. The results
show that the SNe Ia data favor a negative coupling and an
equation of state $w_X<-1$, namely a negatively coupled phantom
dark energy. However, as we know, the supernovae data alone are
not sufficient to constrain dark energy models strictly (see e.g.
the analysis in Ref. \cite{holoSN}). Therefore, to obtain more
tight constraints on dark energy models, one should need
additional data provided by other astronomical observations to be
necessary and useful complements to the SNe data. It has been
demonstrated that some observational quantities irrelevant to
$H_0$ are very suitable to play this role \cite{wangyun}. Such
quantities and data can be found in the probes of CMB and LSS
\cite{holoSN,wangyun,Xia,starob,gyg}. In what follows we perform a
combined analysis of SNe Ia, CMB, and LSS on the constraints of
the NGCG model. We use a $\chi^2$ statistic
\begin{equation}
\chi^2=\chi_{\rm SN}^2+\chi_{\rm CMB}^2+\chi_{\rm LSS}^2~,
\end{equation} where $\chi_{\rm SN}^2$, $\chi_{\rm CMB}^2$ and
$\chi_{\rm LSS}^2$ are contributions from SNe Ia, CMB, and LSS
data, respectively. It is well known that the acceleration of the
Universe is found by the Type Ia supernovae observations, where
the concept of the luminosity distance plays a very important
role. The luminosity distance of a light source is defined in such
a way as to generalize to an expanding and curved space the
inverse-square law of brightness valid in a static Euclidean
space,
\begin{equation}
d_L=\left({{\cal L}\over 4\pi{\cal
F}}\right)^{1/2}=cH_0^{-1}(1+z)\int_0^z{dz'\over E(z')}~,
\end{equation} where ${\cal L}$ is the absolute luminosity which
is a known value for the standard candle SNe Ia, $\cal F$ is the
measured flux. The Hubble distance $cH_0^{-1}=2997.9h^{-1}$ Mpc.
The Type Ia supernova observations directly measure the apparent
magnitude $m$ of a supernova and its red-shift $z$. The apparent
magnitude $m$ is related to the luminosity distance $d_L$ of the
supernova through
\begin{equation}
m(z)=M+5\log_{10}(d_L(z)/{\rm Mpc})+25~,
\end{equation} where $M$ is the absolute magnitude which is
believed to be constant for all Type Ia supernovae. In our
analysis, we take the 157 gold data points listed in Riess et al.
\cite{Riess} which includes recent new 14 high redshift SNe (gold)
data from the HST/GOODS program. The $\chi^2$ function determined
by SNe Ia observations is
\begin{equation}
\chi_{\rm SN}^2=\sum_{i=1}^{157}{[\mu_{\rm obs}(z_i)-\mu_{\rm
th}(z_i)]^2\over \sigma_i^2}~,\label{chisn}
\end{equation} where the extinction-corrected distance moduli
$\mu(z)$ is defined as $\mu(z)=m(z)-M$, and $\sigma_i$ is the
total uncertainty in the observation. Following the Ref.
\cite{IXCDM}, we fix $\Omega_b^0=0.05$ in the computation for
simplification. Hence, the computation is carried out in a
four-dimensional space, for the four parameters $P=(\eta, w_X, h,
\Omega_{dm}^0)$. For the CMB, we use only the measurement of the
CMB shift parameter \cite{cmbshift},
\begin{equation}
{\cal R}=\sqrt{\Omega_m^0}\int_0^{z_{\rm dec}}{dz\over
E(z)}~,\label{R}
\end{equation} where $\Omega_m^0=\Omega_{dm}^0+\Omega_b^0$, and
$z_{\rm dec}=1089$ \cite{wmap}. Note that this quantity is
irrelevant to the parameter $H_0$ such that provides robust
constraint on the dark energy model. The results from CMB data
correspond to ${\cal R}_0=1.716\pm 0.062$ (given by WMAP, CBI,
ACBAR) \cite{wmap,cbi}. We include the CMB data in our analysis by
adding $\chi_{\rm CMB}^2=[({\cal R}-{\cal R}_0)/\sigma_{\cal
R}]^2$ (see e.g. Refs. \cite{wangyun,Xia,starob}), where ${\cal
R}$ is computed by the NGCG model using equation (\ref{R}). The
only large scale structure information we use is the parameter $A$
measured by SDSS \cite{sdssred}, defined by
\begin{equation}
A=\sqrt{\Omega_m^0}E(z_1)^{-1/3}\left[{1\over
z_1}\int_0^{z_1}{dz\over E(z)}\right]^{2/3}~,\label{A}
\end{equation} where $z_1=0.35$. Also, we find that this quantity
is independent of $H_0$ either, thus can provide another robust
constraint on the model. The SDSS gives the measurement data
\cite{sdssred} $A_0=0.469\pm 0.017$. We also include the LSS
constraint in our analysis by adding $\chi_{\rm
LSS}^2=[(A-A_0)/\sigma_A]^2$ (see e.g. Refs. \cite{holoSN,gyg}),
where $A$ is computed by the NGCG model using equation (\ref{A}).
Note that we have chosen to use only the most conservative and
robust information, ${\cal R}$ and $A$, from CMB and LSS
observations. These measurements we use do not depend on the
Hubble constant $H_0$, thus are useful complements to the SNe
data. It is remarkable that the likelihood analysis scheme we
employ here is very economical and efficient due to that it does
not make use of all the information available in CMB and LSS but
can provide fairly good constraints on dark energy models
\cite{holoSN,gyg}.

We now analyze the probability distribution of $\eta$ and $w_X$ in
the NGCG model. The likelihood of these two parameters is
determined by minimizing over the ``nuisance'' parameters
\begin{equation}
{\cal L}(\eta,w_X)=\int dhd\Omega_{dm}^0~ e^{-\chi^2/2}~,
\end{equation} where the integral is over a large enough range of
$h$ and $\Omega_{dm}^0$ to include almost all the probability. We
now compute ${\cal L}(\eta,w_X)$ on a two-dimensional grid spanned
by $\eta$ and $w_X$. The $68.3\%$, $95.4\%$, and $99.7\%$ (namely
1, 2, and 3 $\sigma$) confidence contours consist of points where
the likelihood equals $e^{-2.31/2}$, $e^{-6.18/2}$, and
$e^{-11.83/2}$ of the maximum value of the likelihood,
respectively. Fig.7 shows our main results, the contours of
$1\sigma$, $2\sigma$, and $3\sigma$ confidence levels in the
$w_X-\eta$ plane. The 1 $\sigma$ fit values for the model
parameters are: $w_X=-0.98^{+0.15}_{-0.20}$ and
$\eta=1.06^{+0.20}_{-0.16}$, and the minimum value of $\chi^2$ in
the four dimensional parameter space is: $\chi_{\rm
min}^2=167.29$. We see clearly that the combined analysis of SNe
Ia, CMB, and LSS data provides a fairly tight constraint on the
NGCG model. It is remarkable that the best fit happens at the
vicinity of the cosmological constant, even though $w_X$ is
slightly larger than $-1$ and $\eta$ is mildly larger than 1 (i.e.
$\alpha$ slightly larger than 0). This means that within the
framework of the NGCG model the real form of dark energy at the
maximum probability is the near cosmological constant according to
the joint analysis of SNe+CMB+LSS data. However, the analysis
results still accommodate the existence likelihood of the
``X-matter'' and the interaction between dark energy and dark
matter. In 1 $\sigma$ range, $w_X\in (-1.18, -0.83)$ and
$\alpha\in (-0.1, 0.26)$. This implies that the probabilities of
that dark energy behaves as quintessence-like form and
phantom-like form are roughly equal, and the probabilities that
the energy flow streams from dark energy to dark matter and the
reverse are also roughly equal. One-dimensional likelihood
distribution functions for $w_X$ and $\eta$ are shown in Fig.8 and
Fig.9, respectively. It is very clear that the original Chaplygin
gas model, $\alpha=1$ (or $\eta=2$) and $w_X=-1$, is totally ruled
out by the observational data at $99.7\%$ confidence level. In
addition, it should be pointed out that when we fix $\eta=2$ and
let $w_X$ free, the NGCG model will be identified as the so-called
variable Chaplygin gas (VCG) model proposed in Ref. \cite{Guo1}
(see also Ref. \cite{Guo2}). The joint analysis of SNe+CMB+LSS
also rules out this probability. It is hopeful that the future
precise data will provide more strong evidences to judge whether
the dark energy is the cosmological constant and whether dark
energy and dark matter are in unification.

%%%%%%%%%%%%%%%%%%%%%%%%%%%%%%%%%%%%%%%%%%%%%%%%%%%%%%%%%%%%%%%%$
\vskip.8cm
\begin{figure}
\begin{center}
\leavevmode \epsfbox{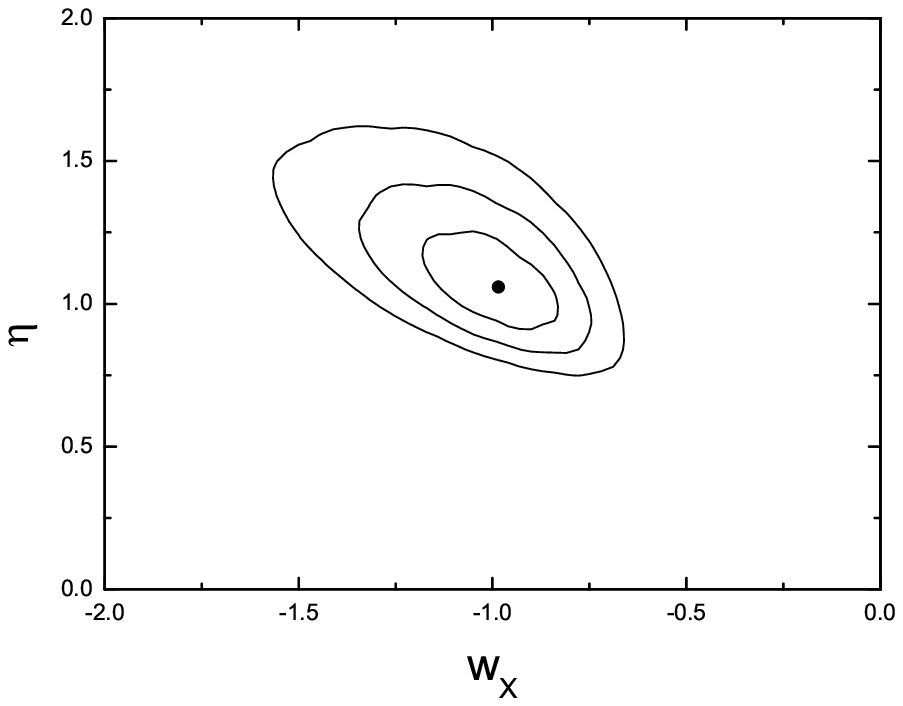} \caption[]{Confidence level
contours of $68.3\%,~95.4\%$ and $99.7\%$ in the $(\eta,w_X)$
plane. The 1 $\sigma$ fit values for the model parameters are:
$w_X=-0.98^{+0.15}_{-0.20}$ and $\eta=1.06^{+0.20}_{-0.16}$, and
the minimum value of $\chi^2$ in the four dimensional parameter
space is: $\chi_{\rm min}^2=167.29$.}
\end{center}
\end{figure}
%%%%%%%%%%%%%%%%%%%%%%%%%%%%%%%%%%%%%%%%%%%%%%%%%%%%%%%%%%%%%%%%%

%%%%%%%%%%%%%%%%%%%%%%%%%%%%%%%%%%%%%%%%%%%%%%%%%%%%%%%%%%%%%%%%$
\vskip.8cm
\begin{figure}
\begin{center}
\leavevmode \epsfbox{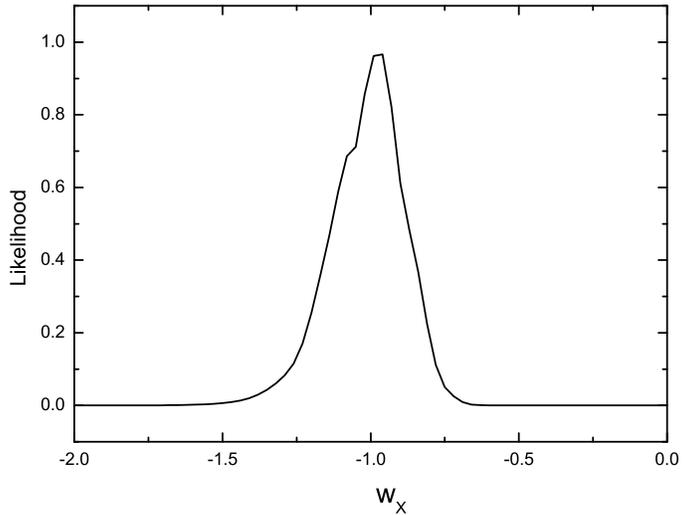} \caption[]{One-dimensional
probability distribution for $w_X$.}
\end{center}
\end{figure}
%%%%%%%%%%%%%%%%%%%%%%%%%%%%%%%%%%%%%%%%%%%%%%%%%%%%%%%%%%%%%%%%%

%%%%%%%%%%%%%%%%%%%%%%%%%%%%%%%%%%%%%%%%%%%%%%%%%%%%%%%%%%%%%%%%$
\vskip.8cm
\begin{figure}
\begin{center}
\leavevmode \epsfbox{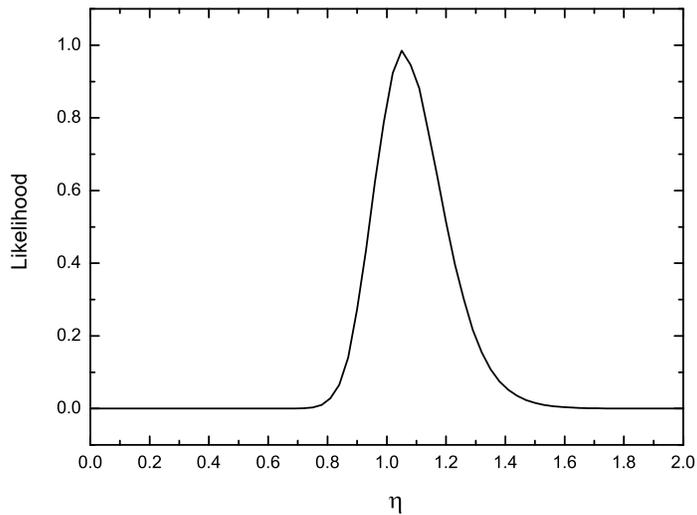} \caption[]{One-dimensional
probability distribution for $\eta$.}
\end{center}
\end{figure}
%%%%%%%%%%%%%%%%%%%%%%%%%%%%%%%%%%%%%%%%%%%%%%%%%%%%%%%%%%%%%%%%%

\section{Concluding remarks}

The Chaplygin gas model is a proposal to describe dark energy and
dark matter as a unified fluid, the Chaplygin gas, characterized
by an exotic equation of state $p=-A/\rho$, where $A$ is a
positive constant. Since this original Chaplygin gas has been
ruled out by the observations, a generalization of the Chaplygin
equation of state $p=-A/\rho^\alpha$ was considered, by
introducing a free parameter $\alpha$. The generalized Chaplygin
gas model is also regarded as a unification of dark energy and
dark matter. The reason is that the GCG behaves as a dust-like
matter at early stage and as a cosmological constant at late
stage. That is to say that the GCG model admits that the Universe
will be dominated by a cosmological constant and thus enter into a
de Sitter phase in the future. However, we can not hitherto affirm
whether the dark energy is a tiny positive cosmological constant.
Therefore, the scheme for unification of dark energy and dark
matter should accommodate other forms of dark energy such as
quintessence-like and phantom-like dark energy. This is the
motivation for us to further generalize the GCG model.

We propose in this paper a new model as a scheme for the
unification of dark energy and dark matter. This new model is a
further extension version of the GCG model, thus dubbed new
generalized Chaplygin gas model. This further generalization is
implemented by introducing another free parameter $w_X$ to make
the constant $A$ in the GCG equation of state become a scale
factor dependent function $\tilde{A}(a)$. In order to implement
the interpolation between a dust dominated Universe and an
X-matter dominated Universe, the unique choice of $\tilde{A}(a)$
is $\tilde{A}(a)=-w_XAa^{-3(1+w_X)(1+\alpha)}$. Through a
two-fluid decomposition, we show that the NGCG model is totally
equivalent to an interacting XCDM parametrization scenario, in
which the interaction between dark energy and dark matter is
characterized by the constant $\alpha$. We discuss the
cosmological consequences led by such an unified dark sectors
model. Furthermore, a statefinder diagnostic is performed on this
scenario and the discrimination between this scenario and other
dark energy models is shown. Finally, a combined analysis of the
data of SNe Ia, CMB, and LSS is used to constrain the parameters
of the NGCG model. The fit result shows that the joint analysis
can provide a considerably tight constraint on the NGCG model.
According to the observational test, the best fit happens at the
vicinity of the $\Lambda$CDM. We hope that the future precise data
will provide more strong evidences to judge whether the dark
energy is the cosmological constant and whether dark energy and
dark matter can be unified into one component. Also, it would be
interesting to investigate the evolution of density perturbations
and the structure formation in the NGCG scenario.

\begin{acknowledgments}
One of us (X.Z.) is grateful to Xiao-Jun Bi, Zhe Chang, Tong Chen,
Zong-Kuan Guo, and Hao Wei for helpful discussions. This work was
supported by the Natural Science Foundation of China (Grant No.
111105035001).
\end{acknowledgments}

%%%%%%%%%%%%%%%%%%%%%%%%%%%%%%%%%%%%%%%%%%%%%%%%%%%%%%%%%%%%%%%%%%%%%%%%%%%%%%

%%%%%%%%%%%%%%%%%%%%%%%%%%%%%%%%%%%%%%%%%%%%%%%%%%%%%%%%%%%%%%%%%%%%%%%%%%%%%%%%%


\begin{thebibliography}{99}
\bibitem{sn}A.~G.~Riess {\it et al.},
%``Observational Evidence from Supernovae for an Accelerating Universe and a
%Cosmological Constant,''
Astron.\ J.\  {\bf 116} (1998) 1009 [astro-ph/9805201];
S.~Perlmutter {\it et al.},
%``Measurements of Omega and Lambda from 42 High-Redshift Supernovae,''
Astrophys.\ J.\  {\bf 517} (1999) 565 [astro-ph/9812133].


\bibitem{wmap}C.~L.~Bennett {\it et al.},
%``First Year Wilkinson Microwave Anisotropy Probe (WMAP) Observations:
%Preliminary Maps and Basic Results,''
Astrophys.\ J.\ Suppl.\  {\bf 148} (2003) 1 [astro-ph/0302207];
D.~N.~Spergel {\it et al.},
%``First Year Wilkinson Microwave Anisotropy Probe (WMAP) Observations:
%Determination of Cosmological Parameters,''
Astrophys.\ J.\ Suppl.\  {\bf 148} (2003) 175 [astro-ph/0302209].

\bibitem{sdss} M.~Tegmark {\it et al.}, Phys.\ Rev.\ D {\bf 69}
(2004) 103501 [astro-ph/0310723]; M.~Tegmark {\it et al.},
Astrophys.\ J.\ {\bf 606} (2004) 702 [astro-ph/0310725].


\bibitem{de}S.~Weinberg,
%``The Cosmological Constant Problem,''
Rev.\ Mod.\ Phys.\  {\bf 61} (1989) 1; S.~M.~Carroll,
%"The Cosmological Constant,"%
Living\ Rev.\ Rel.\ {\bf 4} (2001) 1 [astro-ph/0004075];
P.~J.~E.~Peebles and B.~Ratra,
%``The cosmological constant and dark energy,''
Rev.\ Mod.\ Phys.\  {\bf 75} (2003) 559 [astro-ph/0207347];
T.~Padmanabhan,
%``Cosmological constant: The weight of the vacuum,''
Phys.\ Rept.\  {\bf 380} (2003) 235 [hep-th/0212290].

\bibitem{quin}C.~Wetterich,
%``Cosmology And The Fate Of Dilatation Symmetry,''
Nucl.\ Phys.\ B {\bf 302} (1988) 668; P.~J.~E.~Peebles and
B.~Ratra,
%``Cosmology With A Time Variable Cosmological 'Constant',''
Astrophys.\ J.\  {\bf 325} (1988) L17; R.~R.~Caldwell, R.~Dave and
P.~J.~Steinhardt,
%``Cosmological Imprint of an Energy Component with General
%Equation-of-State,''
Phys.\ Rev.\ Lett.\  {\bf 80} (1998) 1582 [astro-ph/9708069];
I.~Zlatev, L.~Wang and P.~J.~Steinhardt,
% "Quintessence, Cosmic Coincidence, and the Cosmological Constant,"%
Phys.\ Rev.\ Lett.\ {\bf 82} (1999) 896 [astro-ph/9807002];
P.~J.~Steinhardt, L.~Wang and I.~Zlatev,
%``Cosmological tracking solutions,''
Phys.\ Rev.\ D {\bf 59} (1999) 123504 [astro-ph/9812313].

\bibitem{couple} L.~Amendola,
%"Coupled Quintessence,"
Phys.\ Rev.\ D {\bf 62} (2000) 043511 [astro-ph/9908023];
L.~Amendola, D.~Tocchini-Valentini,
%"Stationary dark energy: the present Universe as a global attractor,"
Phys.\ Rev.\ D {\bf 64} (2001) 043509 [astro-ph/0011243];
L.~Amendola, D.~Tocchini-Valentini,
%"Baryon bias and structure formation in an accelerating Universe,"
Phys.\ Rev.\ D {\bf 66} (2002) 043528 [astro-ph/0111535];
L.~Amendola,
%"Acceleration at z>1?"
Mon.\ Not.\ Roy.\ Astron.\ Soc.\ {\bf 342} (2003) 221
[astro-ph/0209494]; M.~Pietroni,
%"Brane Worlds and the Cosmic Coincidence Problem,"
Phys.\ Rev.\ D {\bf 67} (2003) 103523 [hep-ph/0203085];
D.~Comelli, M.~Pietroni, and A.~Riotto,
%"Dark Energy and Dark Matter,"
Phys.\ Lett.\ B {\bf 571} (2003) 115 [hep-ph/0302080]; U.~Franca,
and R.~Rosenfeld,
%"Age constraints and fine tuning in variable-mass particle models,"
Phys.\ Rev.\ D {\bf 69} (2004) 063517 [astro-ph/0308149];
X.~Zhang, Mod. Phys. Lett. A {\bf 20} (2005) 2575
[astro-ph/0503072].


\bibitem{kess}T.~Chiba, T.~Okabe and M.~Yamaguchi,
%``Kinetically driven quintessence,''
Phys.\ Rev.\ D {\bf 62} (2000) 023511 [astro-ph/9912463];
C.~Armendariz-Picon, V.~Mukhanov and P.~J.~Steinhardt,
%``A dynamical solution to the problem of a small cosmological constant  and
%late-time cosmic acceleration,''
Phys.\ Rev.\ Lett.\  {\bf 85} (2000) 4438 [astro-ph/0004134].
C.~Armendariz-Picon, V.~Mukhanov and P.~J.~Steinhardt,
%``Essentials of k-essence,''
Phys.\ Rev.\ D {\bf 63} (2001) 103510 [astro-ph/0006373];
T.~Chiba,
%``Tracking k-essence,''
Phys.\ Rev.\ D {\bf 66} (2002) 063514 [astro-ph/0206298].

\bibitem{xmatter} V. Sahni and A. Starobinsky, Int. J. Mod. Phys. D {\bf9} (2000)
373 [astro-ph/9904398]; V. Sahni, Class. Quant. Grav. {\bf19}
(2002) 3435 [astro-ph/0202076]; L. A. Urena-Lopez and T. Matos,
Phys. Rev. D {\bf62} (2000) 081302 [astro-ph/0003364].



\bibitem{brane} L. Randall and  R. Sundrum, Phys. Rev. Lett. {\bf83}
(1999) 4690- [hep-th/9906064]; T. Shiromizu, K. Maeda, and M.
Sasaki, Phys. Rev. D {\bf62} (2000) 024012 [gr-qc/9910076]; R.
Maartens, D. Wands, B. Bassett, and I. Heard, Phys. Rev. D {\bf62}
(2000) 041301 [hep-ph/9912464]; C. Deffayet, G. Dvali, and G.
Gabadadze, Phys. Rev. D {\bf65} (2002) 044023 [astro-ph/0105068];
C. Deffayet, S. J. Landau, J. Raux, M. Zaldarriaga, and P. Astier,
Phys. Rev. D {\bf66} (2002) 024019 [astro-ph/0201164]; V. Sahni
and Y. Shtanov, JCAP {\bf0311} (2003) 014 [astro-ph/0202346]; H.
Collins and B. Holdom, Phys. Rev. D {\bf62} (2000) 105009
[hep-ph/0003173]; R. Maartens, Living Rev. Rel. {\bf7} (2004) 7
[gr-qc/0312059].

\bibitem{tachyon} G. W. Gibbons, Phys. Lett. B {\bf537} (2002) 1
[hep-th/0204008]; M. Fairbairn and M. H. G. Tytgat, Phys. Lett. B
{\bf546} (2002) 1 [hep-th/0204070];S. Mukohyama, Phys. Rev. D
{\bf66} [hep-th/0204084]; (2002) 024009 A. Frolov, L. Kofman, and
A. Starobinsky, Phys. Lett. B {\bf545} (2002) 8 [hep-th/0204187];
Y.-S. Piao, R.-G. Cai, X. Zhang, and Y.-Z. Zhang, Phys. Rev. D
{\bf66} (2002) 121301 [hep-ph/0207143]; T. Padmanabhan and T. R.
Choudhury, Phys. Rev. D {\bf66} (2002) 081301 [hep-th/0205055]; T.
Padmanabhan, Phys. Rev. D {\bf66} (2002) 021301 [hep-th/0204150].

\bibitem{chaplygin} A. Y. Kamenshchik, U. Moschella, and V.
Pasquier, Phys. Lett. B {\bf511} (2001) 265 [gr-qc/0103004]; N.
Bilic, G. B. Tupper, and R. D. Viollier, Phys. Lett. B {\bf535}
(2002) 17 [astro-ph/0111325]; M. C. Bento, O. Bertolami, and A. A.
Sen, Phys. Rev. D {\bf66} (2002) 043507 [gr-qc/0202064].

\bibitem{gcgtest} M. C. Bento, O. Bertolami, and A. A. Sen, Phys. Rev. D {\bf 67} (2003)
063003 [astro-ph/0210468]; M. C. Bento, O. Bertolami, and A. A.
Sen, Phys. Rev. D {\bf70} (2004) 083519 [astro-ph/0407239];  M. C.
Bento, O. Bertolami, and A. A. Sen, Phys. Lett. B {\bf575} (2003)
172 [astro-ph/0303538]; O. Bertolami, A. A. Sen, S. Sen, and P. T.
Silva, Mon. Not. Roy. Astron. Soc. {\bf353} (2004) 329
[astro-ph/0402387]; M. C. Bento, O. Bertolami, N. M. C. Santos,
and A .A. Sen, Phys. Rev D {\bf71} (2005) 063501
[astro-ph/0412638]; J. S. Alcaniz, D. Jain, and A. Dev, Phys. Rev.
D 67 (2003) 043514 [astro-ph/0210476].

\bibitem{holography} M. Li,
%A Model of Holographic Dark Energy
Phys. Lett. B {\bf603} (2004) 1 [hep-th/0403127]; K. Ke and M. Li,
%Cardy-Verlinde Formula and Holographic Dark Energy
Phys. Lett. B {\bf606} (2005) 173 [hep-th/0407056]; Y. Gong,
%Extended Holographic dark energy
Phys. Rev. D {\bf70} (2004) 064029 [hep-th/0404030]; Y. S. Myung,
%Holographic principle and dark energy
Phys. Lett. B {\bf610} (2005) 18 [hep-th/0412224]; Q. G. Huang and
M. Li,
%The Holographic Dark Energy in a Non-flat Universe
JCAP {\bf0408} (2004) 013 [astro-ph/0404229];
 Q. G. Huang and M. Li,
%Anthropic Principle Favors the Holographic Dark Energy
JCAP {\bf0503} (2005) 001 [hep-th/0410095]; Q. G. Huang and Y.
Gong,
%Supernovae constraints on a holographic dark energy model
JCAP {\bf0408} (2004) 006 [astro-ph/0403590]; Y. Gong, B. Wang and
Y. Z. Zhang, Phys. Rev. D {\bf 72} (2005) 043510
%The Holographic dark energy revisited
[hep-th/0412218]; Z. Chang, F.-Q. Wu, and X. Zhang,
astro-ph/0509531.

\bibitem{holoSN} X. Zhang and F.-Q. Wu, Phys. Rev. D {\bf 72} (2005) 043524
[astro-ph/0506310].

\bibitem{Knop} R. A. Knop et al., Astrophys. J. {\bf598} (2003)
102 [astro-ph/0309368].

\bibitem{Riess} A. G. Riess et al., Astrophys. J. {\bf607} (2004) 665
[astro-ph/0402512].

\bibitem{phantom} R. R. Caldwell, Phys. Lett. B {\bf545} (2002)
23 [astro-ph/9908168].

\bibitem{bigrip} R. R. Caldwell, M. Kamionkowski, and N. N.
Weinberg, Phys. Rev. Lett. {\bf91} (2003) 071301
[astro-ph/0302506].

\bibitem{gcgbrane} N. Bilic, G. B. Tupper, and R. D. Viollier,
astro-ph/0207423.

\bibitem{susy} R. Jackiw, ``A Particle Field Theorist's Lectures on Supersymmetric,
Non-Abelian Fluid Mechanics and d-Branes'', physics/0010042.

\bibitem{IXCDM} E. Majerotto, D. Sapone, and L. Amendola,
astro-ph/0410543.

\bibitem{Cai} R.-G. Cai and A. Wang, JCAP {\bf0503} (2005) 002
[hep-th/0411025].

\bibitem{sahni} V.~Sahni, T.~D.~Saini, A.~A.~Starobinsky and
U.~Alam,
%Statefinder -- a new geometrical diagnostic of dark energy
JETP Lett. {\bf 77} (2003) 201 [astro-ph/0201498].

\bibitem{alam} U.~Alam, V.~Sahni, T.~D.~Saini and
A.~A.~Starobinsky,
%Exploring the Expanding Universe and Dark Energy using the Statefinder Diagnostic
Mon.\ Not.\ Roy.\ ast.\ Soc.\ {\bf 344} (2003) 1057
[astro-ph/0303009].

\bibitem{quintomsr} P. Wu and H. Yu, gr-qc/0509036.

\bibitem{gorini} V.~Gorini, A.~Kamenshchik and U.~Moschella,
%Can the Chaplygin gas be a plausible model for dark energy?
Phys.\ Rev.\ D {\bf 67} (2003) 063509 [astro-ph/0209395].

\bibitem{holosr} X. Zhang, Int. J. Mod. Phys. D 14 (2005) 1597
[astro-ph/0504586].

\bibitem{zimdahl} W.~Zimdahl and D.~Pavon,
%Statefinder parameters for interacting dark energy
Gen.\ Rel.\ Grav. {\bf 36} (2004) 1483 [gr-qc/0311067].

\bibitem{zx} X.~Zhang,
%"Statefinder diagnostic for coupled quintessence,"
Phys.\ Lett.\ B {\bf 611} (2005) 1 [astro-ph/0503075].

\bibitem{wangyun}  Y. Wang, and P. Mukherjee, Astrophys. J. {\bf606}
(2004) 654 [astro-ph/0312192]; Y. Wang, and M. Tegmark, Phys. Rev.
Lett. {\bf92} (2004) 241302 [astro-ph/0403292].

\bibitem{Xia} J.-Q. Xia, B. Feng, and X. Zhang, Mod. Phys. Lett. A {\bf20} (2005)
2409 [astro-ph/0411501].

\bibitem{starob} U. Alam, V. Sahni, and A. A. Starobinsky, JCAP
0406 (2004) 008 [astro-ph/0403687].


\bibitem{gyg} Y. Gong, and Y.-Z. Zhang, Phys. Rev. D {\bf72} (2005) 043518
[astro-ph/0502262].

\bibitem{cmbshift} J. R. Bond, G. Efstathiou, and M. Tegmark,
Mon. Not. Roy. Astron. Soc. {\bf 291} (1997) L33
[astro-ph/9702100];  A. Melchiorri, L. Mersini, C. J. Odman, and
M. Trodden, Phys. Rev. D {\bf68} (2003) 043509 [astro-ph/0211522];
C. J. Odman, A. Melchiorri, M. P. Hobson, and A. N. Lasenby, Phys.
Rev. D {\bf67} (2003) 083511 [astro-ph/0207286].

\bibitem{cbi} CBI Collaboration, T. J. Pearson et al., Astrophys. J. {\bf591}
(2003) 556 [astro-ph/0205388]; ACBAR Collaboration, C. L. Kuo et
al., Astrophys. J. {\bf600} (2004) 32 [astro-ph/0212289].

\bibitem{sdssred} D. J. Eisenstein et al., astro-ph/0501171.

\bibitem{Guo1} Z.-K. Guo and Y.-Z. Zhang, astro-ph/0506091.

\bibitem{Guo2} Z.-K. Guo and Y.-Z. Zhang, astro-ph/0509790.



























\end{thebibliography}
\end{document}